\documentclass[fleqn,usenatbib,useAMS]{mnras}

\usepackage{mathptmx}

\usepackage[T1]{fontenc}

\usepackage{graphicx}	
\usepackage{amsmath}	
\usepackage{amssymb}	

\newcommand{\etal}{et al.}
\newcommand{\hst}{\textit{HST}}
\newcommand\Mbh{M\textsubscript{\text{bh}}}
\newcommand\zobs{z\textsubscript{\text{obs}}}
\newcommand\fgout{f\textsubscript{\text{g,out}}}
\newcommand\Rout{R\textsubscript{\text{out}}}
\newcommand\Rin{R\textsubscript{\text{in}}}

\newcommand\Mstardot{\dot{M}_{*}}
\newcommand\Teff{T\textsubscript{\text{eff}}}

\newcommand\tdyn{\tau\textsubscript{\text{dyn}}}
\newcommand\Omegaout{\Omega\textsubscript{\text{out}}}
\newcommand\Mstardotavg{\langle\Mstardot\rangle}
\newcommand\reff{R\textsubscript{\text{1/2}}}

\title[The remnants of nuclear starburst discs]
  {The stellar remnants of high redshift nuclear starburst discs: a
    potential origin for nuclear star clusters?}
\author[R.\ Gohil, D.\ R.\ Ballantyne and G.\ Li]
  {R.~Gohil\thanks{raj.gohil07@gmail.com}, D.~R.~Ballantyne and G.~Li\\Center
    for Relativistic Astrophysics, School of Physics, Georgia
    Institute of Technology, 837 State Street, Atlanta, GA 30332-0430,
    USA}

\date{Accepted XXX. Received YYY; in original form ZZZ}

\pubyear{2019}

\begin{document}
\label{firstpage}
\pagerange{\pageref{firstpage}--\pageref{lastpage}}
\maketitle

\begin{abstract}
Nuclear starburst discs (NSDs) are very compact star-forming regions in
the centers of galaxies that have been studied as a possible origin for
the absorbing gas around a central active galactic nucleus. NSDs
may be most relevant at
$z\sim 1$ when obscured accretion onto supermassive black holes
(SMBHs) is common. This paper describes the characteristics of the stellar
remnants of NSDs at $z=0.01$, taking into account the evolution from
$z=1$. Using a stellar synthesis model, the colours, masses, and
luminosities of the stellar remnants are computed for a suite of 192 two-dimensional NSD models. These properties are compared to
observations of local nuclear star clusters (NSCs), and a good match
is found between the predicted and observed properties. Dynamical effects will likely cause the
final remnant to be a rotating, nearly spherical distribution. In addition,
$\approx 20$\% of the NSD remnants have
half-light radii $\la 10$~pc, consistent with NSCs hosted in both late-type
and early-type galaxies, and all the remnants follow similar size-luminosity
relationships as observed in nearby NSCs. NSDs require the presence of a central
SMBH and the most massive and compact stellar remnants are associated with the most massive SMBHs, although stellar clusters with a
variety of sizes can be produced by all considered SMBH masses.  Overall,
NSDs at $z\sim 1$ appear to be a promising origin for the $\gg 1$~Gyr
NSC population in early- and late-type galaxies with large SMBHs.  
\end{abstract}


\section{Introduction}
\label{sect:intro}
Nuclear starburst discs (NSDs) --- extremely compact,
Eddington-limited starbursts in the cores of galaxies --- have been
proposed as a potential method for fueling and obscuring active
galactic nuclei (AGNs; \citealt{ball08,gb17}). A key prediction of both
1-dimensional (1D) and 2-dimensional (2D) NSD models is the presence
of a pc-scale burst of star-formation caused by the loss of dust
opacity which inflates the disc atmosphere and results in significant
obscuration along most lines-of-sight to the central AGN \citep*{tqm05,ball08,gb17}. This
pc-scale starburst is more likely to occur when galaxies have high gas fractions, and
therefore NSDs are a possible candidate for obscuring AGNs at $z\sim
1$ and influencing the shape of the X-ray Background \citep{gb18}.

If NSDs are obscuring AGNs at high redshifts, then the
remnants of these discs may be detectable in the nuclei of local
galaxies. Interestingly, over the last several years \textit{Hubble Space Telescope} (\hst)
observations have revealed that nuclear star clusters
(NSCs) are found at the center of most ($>70$\%) nearby early
and late-type galaxies
\citep[e.g.,][]{car97,matthews99,boker02,cote06,seth08,gb14,car15,geo16}. The effective radii (i.e., the half-light
radii) of many NSCs is smaller than a few tens of pc
\citep[e.g.,][]{boker04,cote06,turner12,geo16}, and they show
evidence for a complex star-formation history 
with the presence of multiple generations of stars
\citep[e.g.,][]{walcher05,rossa06,seth06,seth10,lyubenova13,kach18}, including an
older population with ages $\gg 1$~Gyr. In addition, NSCs are also observed to be
rotating \citep{seth08b}, and many of them co-exist with
a central SMBH \citep[e.g.,][]{seth08,gs09,geo16} in their host galaxies. Two broad class of mechanisms have been suggested for the formation of
NSCs. The first focuses on the migration of dense clusters via dynamical friction
\citep[e.g.,][]{andersen08,antonini13,arca14}, while the second one
considers in situ nuclear star-formation from gas inflows
\citep[e.g.,][]{milo04,seth06}. Given the wide range of NSC
properties observed in both early- and late-type galaxies, both of
these mechanisms are likely relevant and operating with various
efficiencies in different galaxies \citep[e.g.,][]{abs15,ger16}.

A very compact ($\sim 10$~pc), rotating, multi-Gyr-old stellar population
associated with a SMBH is very suggestive as a potential remnant
from a NSD that was obscuring an AGN at $z \sim 1$. However, to test
this idea the stellar content of a NSD must be evolved to the present
day to compare with observations of nearby NSCs. This paper presents
the result of such an experiment. The next section provides a brief
overview of the NSD models, and then Section~3 describes the procedure
to compute various properties of the stellar remnants of NSDs. The predicted properties of the
resulting star clusters are presented in Section~4, where a comparison to
observed NSCs is also performed. Finally, Section~5 contains a short
discussion and a summary of the results. A standard $\Lambda$CDM
model cosmology is assumed with $H_0=70$~km s$^{-1}$ Mpc$^{-1}$,
$\Omega_{\Lambda}=0.72$, and $\Omega_m=0.28$ \citep{hinshaw13}. 

\section{Overview of Nuclear Starburst Disc Models}
\label{sect:nsds}
The 1D theory of NSDs was first presented by \citet{tqm05}, and
was extended to 2D by \citet{gb17}. Those papers present a complete
description of the model, but a brief overview of the fundamental
properties and assumptions of the 2D model is presented here.

The 2D NSD models used in this paper are time-independent, equilibrium
calculations of a Keplerian gas disc orbiting in the potential of a
SMBH and galactic bulge. The disc midplane is always assumed to be critically
unstable to star-formation (i.e., the Toomre $Q$ parameter is very
close to one), and the radiation pressure of starlight on dust is
fundamental to the vertical support of the disc. The calculation does
not compute the formation of individual stars, but rather determines the
radial and vertical structure of the NSD by finding the radial
distribution of the star-formation rate that leads to hydrostatic, thermal
and energy balance.

An interesting aspect of the NSD theory is that gas is assumed to slowly accrete through
the disc with a radial velocity $mc_s$, where $c_s$ is the local sound
speed and $m$ is a constant Mach number. This results in a constant
competition between star-formation and accretion at every radius. If
gas reaches a radius $\sim$pc from the SMBH, then the temperature of
the gas may be high enough to sublimate dust. This results
in a large burst of star-formation as well as a vertical expansion
of an atmosphere. The vertically inflated atmosphere is supported by
the radiation pressure from the bursty star-formation
\citep{tqm05,gb17}. \cite{gb17} computed the
two-dimensional structure of 192 NSDs across its input parameter space
and showed that 52\% of these models possess the starburst phenomenon
at the parsec-scale.

\begin{figure}
  \includegraphics[width=0.5\textwidth]{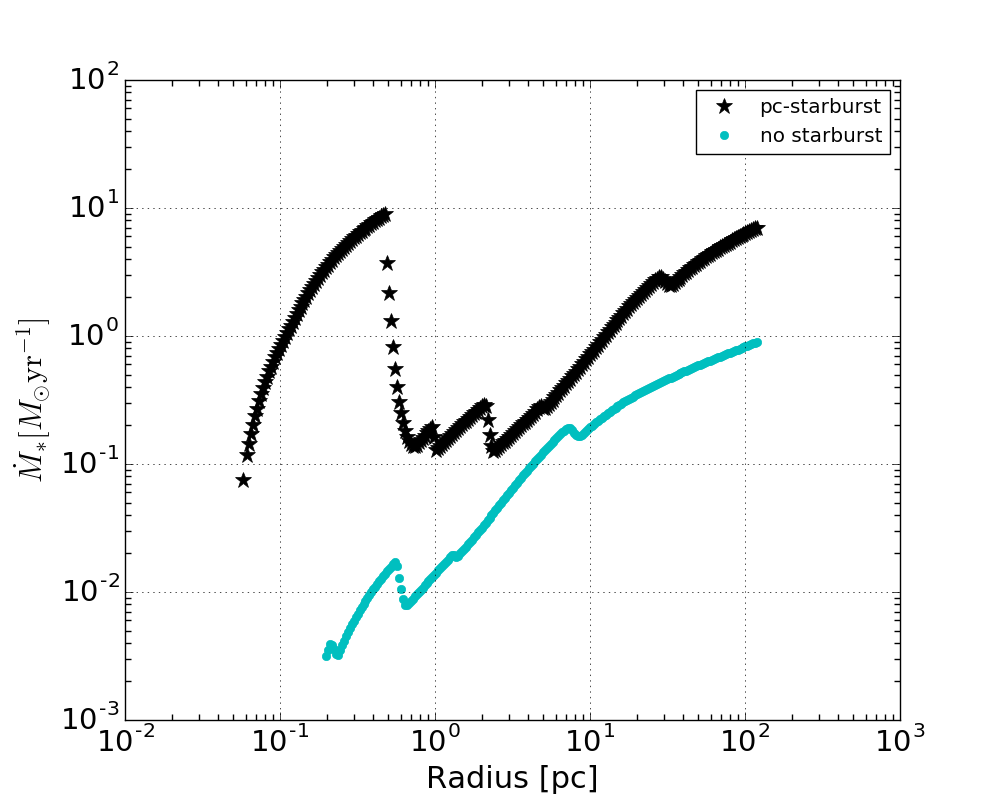}
\caption{Two examples of the equilibrium SFR distributions found in
  the sample of 192 NSD models computed by \citet{gb17}. The black stars show the SFR
  profile for a model ($\Mbh=10^7$~M$_{\odot}$, $\Rout=120$~pc,
  $\fgout=0.6$, $m=0.5$) that produces a pc-scale starburst due to the
  loss of dust opacity at the inner edge of the disc, while the cyan
  circles plot the profile from a different NSD
  ($\Mbh=10^7$~M$_{\odot}$, $\Rout=120$~pc, $\fgout=0.2$, $m=0.5$)
  where most of the  star-formation occurs in the outer disc. See
  \citet{tqm05} and \citet{gb17} for more details.}
\label{fig:sfrs}
\end{figure}

The structure of an individual NSD depends on four input
parameters: the disc size ($\Rout$), the gas fraction at the outer radius
($\fgout$), the Mach number ($m$), and the SMBH mass
($\Mbh$). The distribution of parameters used in the \citet{gb17}
model set is shown in Table~\ref{table:ips}, and were not chosen nor
optimised for describing local NSCs. Therefore, the
NSD remnants derived from this parameter set will provide an
unbiased sample to compare against the observed NSC properties.
\begin{table}
\centering
\caption{All the bins of input parameters for NSDs modeling are
  listed: $\Mbh$, $\Rout$, $\fgout$, and $m$. A total of 192 models
  are computed with all the possible combination of these bins \citep{gb17}.} 
\label{table:ips}
\begin{tabular}{c | c c c c}
log($\Mbh/$M$_{\odot}$) & $\Rout$(pc) & $\fgout$ & $m$ & \\\hline
6.5      & 240  & 0.2   & 0.1  &  \\
7.0      & 180  & 0.4   & 0.3  &  \\
7.5      & 120  & 0.6   & 0.5  &  \\
8.0      & 60   & 0.8   &  --  &
\end{tabular}
\end{table}
Each of these 192 models calculated from the different combination of
parameters yields a radial distribution of
star-formation rates (SFR) needed to vertically support the NSD. The
distribution of SFRs can vary widely among the models depending on
the input parameters. Two examples of the derived equilibrium SFRs are
shown in Figure~\ref{fig:sfrs}.
The profile with the black stars exhibits a pc-scale burst of
star-formation resulting from the loss of dust opacity as described
above, while the SFR distribution shown with the cyan circles has most
of the star-formation at the outer edge of the disc. In general, all
of the 192 NSD models fall into one of the two profiles shown in
Fig.~\ref{fig:sfrs}, but with significant variations in the magnitude
of the SFRs \citep[e.g.,][]{ball08,bai13}. These 192 time-independent SFR profiles derived
from the NSD models are then used to derive the predicted properties
of the relic star cluster long after the end of the starburst.

\section{Calculation of Properties of Remnant Star Cluster}
\label{sect:calc}
\subsection{Timescales and Half-light Radius}
\label{sub:timescales}
As the SFR profiles computed by \citet{gb17} are time-independent, a
characteristic timescale must be identified in order to predict the
final star cluster properties. Here, we use the dynamical time of the NSD, $\tdyn$, which appears
to explain observed star-formation efficiencies in both low-$z$ and
high-$z$ environments
\citep[e.g.,][]{silk97,elmegreen97,kenn98,genzel10}. The dynamical
time is computed for all 192 NSD models by
\begin{equation}
  \label{eq:tdyn}
  \tdyn\equiv \frac{1}{\Omegaout},
\end{equation}
where $\Omegaout$ is the Keplerian angular frequency at $\Rout$.

If the NSD structure does not change significantly over $\tdyn$,
then the final distribution of light in the remnant cluster will follow
the SFR profile predicted by the NSD theory (e.g.,
Fig.~\ref{fig:sfrs}). Therefore, the half-light radius of the final remnant, $\reff$ (also known as the effective
radius), can be predicted directly from the NSD model. As described by
\citet{tqm05} and \citet{gb17}, the total radiative flux due to
star-formation is known at each radius in a NSD, so the bolometric
spectrum can be computed by assuming each annulus emits as a blackbody
with an effective temperature $\Teff(r)$. Then $\reff$ for
each NSD is given by the following equation:
\begin{equation}
  \label{eq:reff}
\int_{\Rin}^{\reff}\int_{\lambda_\text{min}}^{\lambda_\text{max}}f(\lambda,r)drd\lambda=\frac{1}{2}\int_{\Rin}^{\Rout}\int_{\lambda_\text{min}}^{\lambda_\text{max}}f(\lambda,r)drd\lambda,
\end{equation}
where
\begin{equation}
f(r)=\frac{4\pi^2 hc^2r}{\lambda^5\exp{\Big[\frac{hc}{\lambda k_B\Teff(r)}}-1\Big]}
\end{equation}
and $\lambda_\text{min}$ and $\lambda_\text{max}$ are
computed based on the maximum and minimum $\Teff(r)$. This bolometric $\reff$ can then be compared to measurements of
NSCs in the local Universe since the observed values depend weakly
on wavelength \citep[e.g.,][]{gb14}.

\begin{figure}
\includegraphics[width=0.49\textwidth]{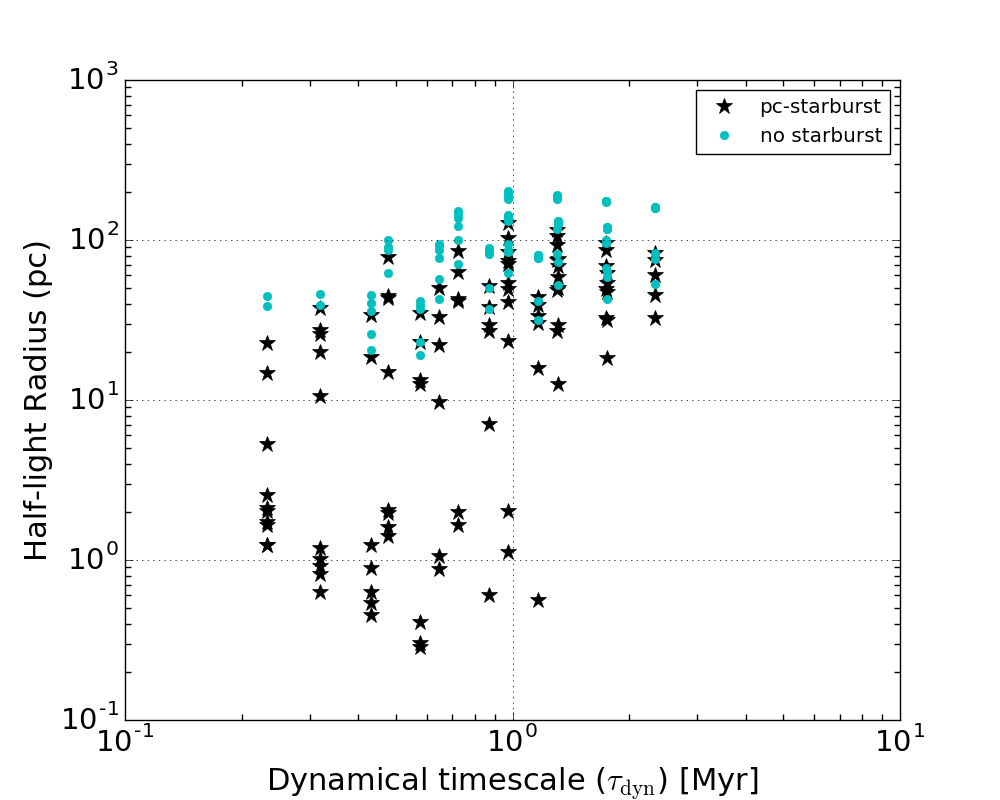}
\centering
\caption{The half-light radius of NSDs computed from
  Eq.~\ref{eq:reff} plotted against the dynamical timescale
  ($\tdyn$; Eq.~\ref{eq:tdyn}). The black stars indicate NSDs that
  produce pc-scale starbursts, while the cyan circles are NSDs that do
  not produce such bursts. The half-light
  radii of the NSDs span a range from $\approx 0.3$~pc to $\approx
  200$~pc (the maximum radius considered by the model suite is
  $\Rout=240$~pc). All the NSD exhibit
  $\tdyn \la 3$~Myrs, and a wide range of $\reff$ are found at $\tdyn
  \la$ 1~Myrs.} 
\label{fig:reff}
\end{figure}

Figure~\ref{fig:reff} plots $\reff$ and $\tdyn$ for all 192
NSD models considered here, including those with (black stars) and
without (cyan circles) the pc-scale burst of star-formation.
The figures show that the NSDs without pc-scale starbursts all have
$\reff \ga 20$~pc, but that the ones with the small scale bursts are
clustered in two groups. There is one group of black stars with
intermediate sizes ($\sim$10s of pc), and a second group with $\reff
\sim 1$~pc. The first group are models where there is roughly equal
amounts of star-formation in the outer disc and in the pc-scale burst,
leading to a $\reff$ that is roughly in the middle of the disc. The
NSDs with the small $\reff$ are ones where the SFR in the pc-scale
burst overwhelms all other star-formation in the disc. The dynamical
times are $ < 3$~Myrs for all models and are spread in vertical stripes because the Keplerian frequency of a disc at the
outer radius depends only on $\Rout$ and $\Mbh$. For a fixed pair of
$\Rout$ and $\Mbh$, there are 12 models exhibiting the same $\tdyn$
due to degeneracy into other parameters. Interestingly,
the plot shows that the NSDs with pc-scale bursts
are found at all $\tdyn$, with a preference for the small $\reff$ NSDs
to be found at the smaller $\tdyn$, corresponding to lower $\Mbh$ and/or smaller disc sizes.

\begin{figure}
  \includegraphics[width=0.50\textwidth]{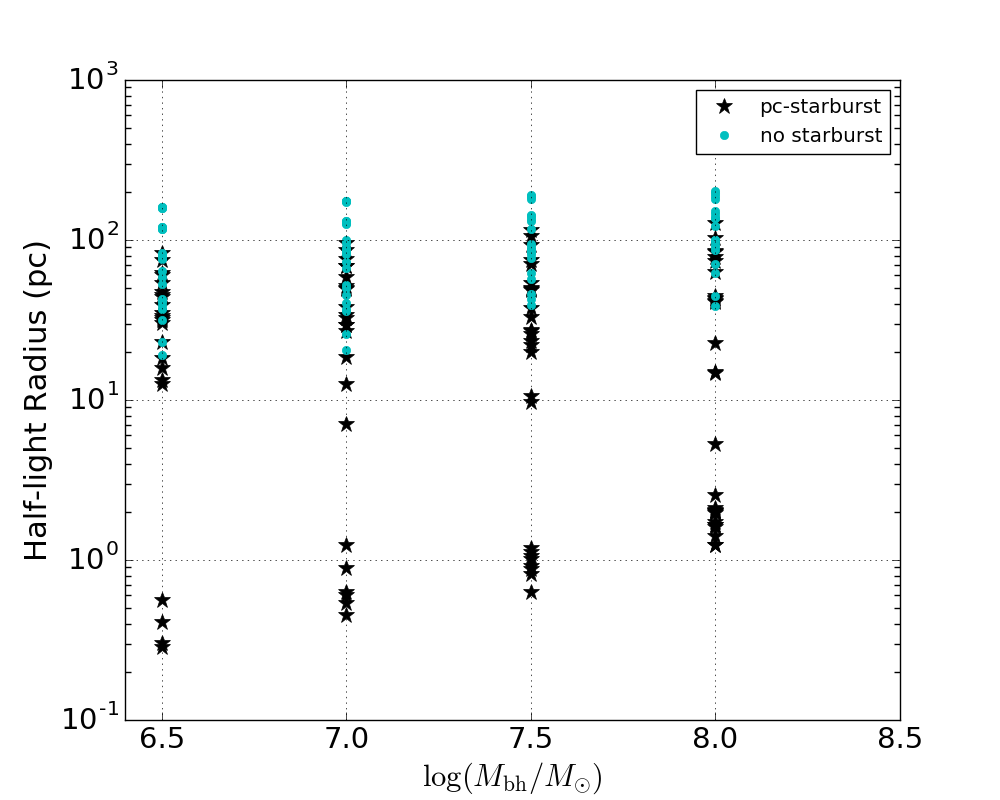}
  \caption{The half-light radius, $\reff$, of NSDs plotted for the
    four values of $\Mbh$ used in our model suite
    (Table~\ref{table:ips}). As in the other figures, the black stars
    indicate NSD models that produce pc-scale starbursts. For the group
    of models with $\reff \la 10$~pc, there is a mild correlation
    between $\reff$ and $\Mbh$, indicating that the most compact
    clusters will be associated with the smallest black
    holes. However, compact clusters are more common for the higher
    black hole masses with a total of 15 NSDs with $\reff < 10$~pc at $\Mbh
  =10^8$~$M_{\odot}$, as compared to 9 at $\Mbh
  =10^{7.5}$~$M_{\odot}$, 7 at $\Mbh =10^7$~$M_{\odot}$, and 4 at $\Mbh
  =10^{6.5}$~$M_{\odot}$.}
  \label{fig:reffmbh}
\end{figure}

To help determine if a smaller $\Mbh$ or $\Rout$ is more important in
producing the pc-scale bursts, Figure~\ref{fig:reffmbh} 
plots $\reff$ versus the $\Mbh$ at the center of the NSD.
Compact NSDs with $\reff < 10$~pc are seen at all $\Mbh$, but are actually
more common at larger $\Mbh$ with 15 at $\Mbh =10^8$~$M_{\odot}$, as
compared to only 4 at $\Mbh =10^{6.5}$~$M_{\odot}$. This implies that a
larger $\Mbh$ is a stronger factor than $\Rout$ in producing the most powerful
pc-scale bursts, in agreement with the results of the 1D calculations
\citep{ball08}. More massive black holes cause hotter and denser
NSDs which more readily lead to dust sublimation and a small scale
burst of star-formation. However, Fig.~\ref{fig:reffmbh} also shows that the
smallest $\reff$ are found in NSDs around the lowest mass SMBHs, so
the size of the stellar remnant will be weakly correlated with $\Mbh$.

\subsection{Mass, Luminosity and Colours}
\label{sub:mass}
The predicted mass and spectral properties of the remnant cluster at
low redshift can only be
calculated by assuming a star-formation history for the NSD models and
evolving the stellar population in time. As before, the equilibrium
radial structure computed for each NSD is assumed to be fixed during
its lifetime. The radially-dependent SFR (e.g., Fig.~\ref{fig:sfrs})
is then averaged to yield $\Mstardotavg$ for each model, i.e.,
\begin{equation}
\Mstardotavg =  {\int_{\Rin}^{\Rout} \Mstardot(r)2\pi r
  dr \over \int_{\Rin}^{\Rout} 2\pi r dr}.
\label{eq:avgsrf}
\end{equation}
The star-formation history of each NSD model is then
assumed to be a narrow Gaussian in time with a lifetime $\tdyn$:
\begin{equation}
  \label{eq:gauss}
\Mstardot(t)=A\exp\Bigg[-\Bigg(\frac{(t-\mu)}{\sqrt2\sigma}\Bigg)^2\Bigg],
\end{equation}
where
$A=\Mstardotavg\tdyn/\sigma\sqrt2\pi$, $\mu=t_0+\tdyn/2$, and
$\sigma=\tdyn/5$. The normalization constant ($A$) is chosen so
that the stellar mass formed in the NSDs is equal to
$\Mstardotavg \tdyn$. The time $t_0$ corresponds to the age of the
Universe at $z=1$ (i.e., $5.9$~Gyrs), when such NSDs may be common due
to larger galaxy gas fractions \citep{gb18}.

The star cluster formed by this star-formation history is then evolved forward in time
using \textsc{Python-FSPS}, the
Python wrapper of the Flexible Stellar Population Synthesis code (\textsc{FSPS}
v3.0) developed by \citet{conroy09} and \citet{conroy10}. All
\textsc{FSPS} calculations assume solar metallicity and a \citet{kroupa01}
initial mass function. The \cite{kriek13} model is used for the
attenuation by dust around old stars. The characteristics of the
remnant star clusters are computed assuming an observed redshift
$\zobs=0.01$, giving an age of the stellar population of $ 7.7$~Gyr\footnote{The
  properties of the remnant cluster show negligible differences for
  other starting redshifts (e.g., $z=0.8$ or $1.5$) that yield final
  cluster ages $\gg 1$~Gyr.}. Each \textsc{FSPS} calculation yields
the survived stellar mass at $z=0.01$, $M_{\ast}$, and absolute magnitudes in a
number of \hst\ and Johnson filters.

\subsection{Dynamical Evolution of $\reff$}
\label{sub:dynamics}
The calculation of $\reff$ described above (Eq.~\ref{eq:reff}) is
based on the radial distribution of stars in each model NSD. It is
possible that dynamical effects could significantly alter this radial
distribution in the $\approx 8$~Gyrs between the NSD event and
$z=0.01$. Therefore, it is important to consider how dynamics within the
remnant stellar discs may impact the values of $\reff$ before
comparing them to observed NSCs at low redshifts.

The stellar cluster produced by a NSD will be a rotating disc spanning
from $\sim 0.1$~pc to $\sim 100$--$200$~pc from the galactic
center. The dynamical effects relevant to stars in the disc are
nicely summarized by \citet{kt11} in their Figure~1. It shows that
a number of dynamical processes have timescales below $\sim 8$~Gyrs
for discs with these sizes, including standard two-body
relaxation, disc eccentricity relaxation, Newtonian precession,
relativistic precession, vector resonant relaxation (VRR), and warps
by a massive perturber (e.g., a molecular torus). With larger SMBH
masses ($\sim 10^8M_{\odot}$), Lense-Thirring precession timescale can
also be shorter than $\sim 8$~Gyrs in the inner edge of the disc. 

However, most of these dynamical processes do not change the radial
distribution of the stars significantly. Specifically, Newtonian
precession and relativistic precession only change the pericenter
orientation of the stellar orbits in their orbital planes. Similarly,
VRR, Lense-Thirring precession and the warp of the disc redistribute
orbital plane orientations of stars in the disc, and have only a
minimal effect on the semi-major axes and the value of $\reff$. In
particular, a recent numerical study of VRR in stellar discs by
\citet{sk18} showed  an interesting mass dependence in the final
stellar distribution. Their simulations found that massive young stars
were in a warped disc, but the less massive older stars were in
a rotating, mostly spherical distribution. If VRR is important for the
NSD remnants, then these results imply that at $z=0.01$ the stellar
cluster will more closely resemble a rotating spheroidal system. 

The effects of non-resonant relaxation (standard two-body relaxation and disc eccentricity relaxation) on stellar discs have been investigated numerically
by \citet{sh14}. These authors find that the relaxation processes
scatter stars both inwards and outwards, which causes peaks in the
semi-major axis distribution to be smoothed out to both higher and
lower values. This numerical result can be rescaled for different SMBH masses and disc sizes, as noted by \citet{sh14}. Therefore, even in
cases where there is a peak in stellar density, such as in NSDs with
$\reff \la 10$~pc, there will be no systematic shift in stellar
density over time.

The stellar discs produced by NSDs may also be subject to spiral-arm
instabilities \citep[e.g.,][]{lg99} at different points in its
evolution. These instabilities cause mixing of stellar orbits as they
pass through the gas, but, as with two-body relaxation, roughly equal
numbers of stars are scattered inwards as outwards and there is no
significant change in the surface density profile of the disc \citep{sb02}.

From the above discussion we conclude that the values of $\reff$
computed by eq.~\ref{eq:reff} will be an adequate estimate of the
effective radii of the stellar remnants of NSDs at $z=0.01$ and can be
compared to the measurements of nearby NSCs. In addition, the effects
of VRR will likely produce a rotating, quasi-spherical remnant at low
redshift. 

\section{Comparison to Low Redshift Nuclear Star Clusters}
\label{sect:nscs}
\begin{figure}
\includegraphics[width=0.5\textwidth]{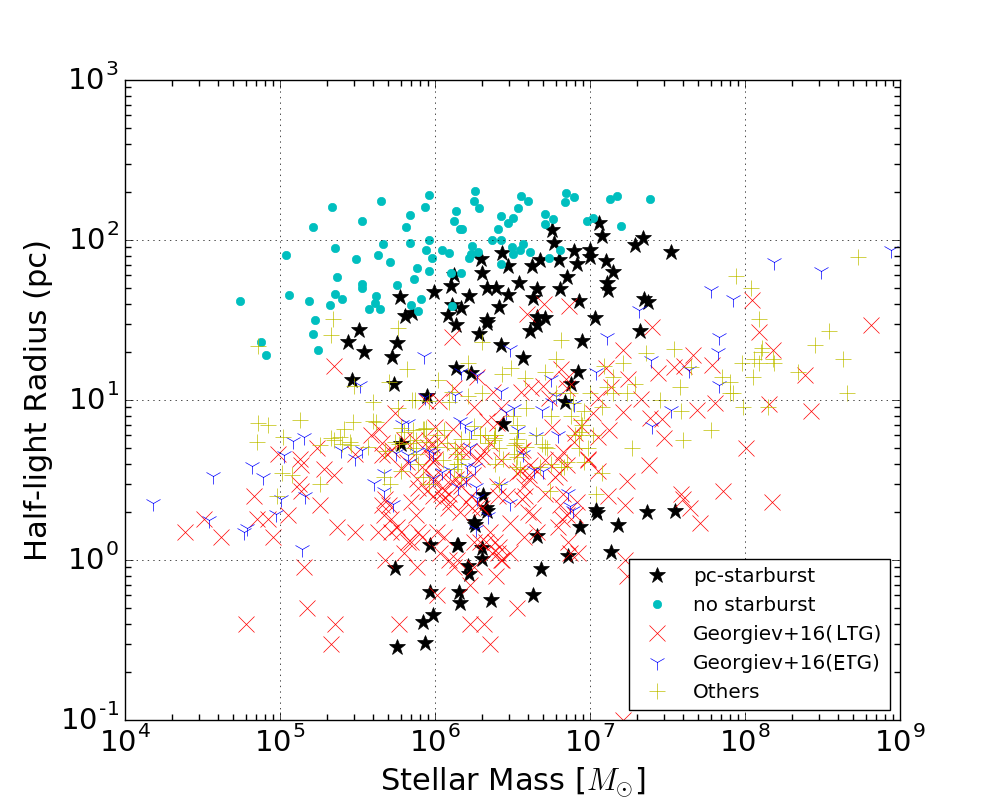}
\centering
\caption{The predicted half-light radius, $\reff$, and remnant mass,
  $M_{\ast}$, calculated from the 192 NSD models as described in
  Sect.~\ref{sect:calc}. The NSDs are assumed to exist at $z=1$ and
  the remnant mass is calculated at $z=0.01$. As in
  Fig.~\ref{fig:reff}, the remnant clusters are separated based on if
  the originating NSD did (black stars) or did not (cyan circles) have
  a pc-scale starburst. Also included in the figure are measurements
  of low redshift NSCs in late-type galaxies (LTG; red `X's)
  and early-type galaxies (ETG; blue `Y's) from the sample of
  \citet{geo16}. The yellow `+'s are data from a heterogeneous sample of
NSCs compiled from the literature
\citep{car97,carollo98,boker04,seth06,seth10,leigh12}.} 
\label{fig:reffmass}
\end{figure}

In this section the computed properties of the NSD remnants at
$z=0.01$ are compared to observations of NSCs observed by \hst\ in the
local Universe. For reference, Table~\ref{table:medians} provides
the median $M_{\ast}$, $\reff$, $\tdyn$, and absolute V-band
magnitude, $M_V$, for four different samples of the NSD models.
\begin{table*}
\centering
\caption{Median properties of the 192 NSDs and their remnant star
  clusters. The 192 models are separated into 4 samples: (i) all
  models, (ii) those without
pc-scale starbursts, (iii) those with pc-scale
bursts and $\reff > 10$~pc, and (iv) those with pc-scale bursts and
$\reff < 10$~pc. The columns are the number of models in the sample,
the dynamical time (Eq.~\ref{eq:tdyn}), the half-light radius (Eq.~\ref{eq:reff}), the
remnant mass at $z=0.01$ and its V-band absolute magnitude.} 
\label{table:medians}
\begin{tabular}{c | c c c c c}
Sample & Number & $\tdyn$(Myr) & $\reff$(pc) & $\log(M_{\ast}/M_{\odot}$) &
                                                                         $M_V$(mag)\\\hline
  All models & 192 & 0.92 & 50 & 6.3 & $-9.6$ \\
No pc-scale burst & 88  & 0.98  & 87 & 6.1  & $-9.1$ \\
pc-scale burst ($\reff > 10$~pc) & 69  & 0.98 & 43 & 6.6 & $-10.3$ \\
pc-scale burst ($\reff < 10$~pc) & 35 & 0.4 & 1.2 & 6.3 & $-9.7$ \\
\end{tabular}
\end{table*}
We begin in Figure~\ref{fig:reffmass} by comparing
$\reff$ and $M_{\ast}$ computed from the NSD models to a large number
of NSC observations collected from the literature
\citep{car97,carollo98,boker04,seth06,seth10,leigh12,geo16}.  
The majority of the NSC masses are computed photometrically using
mass-to-light ratios \citep[e.g.,][]{geo16}, which could lead to an
underestimate of the mass if a young stellar population was dominating
the observed flux. However, dynamical NSC mass measurements
\citep[e.g.,][]{seth08,nguyen18} agree with the photometric ones to
within a factor of about $2$, indicating that there is no large systematic error with the
photometric masses. Therefore, the NSC masses plotted in
Fig.~\ref{fig:reffmass} are likely good estimates of the mass of the
older stellar population in NSCs, and the comparison to the NSD
remnant masses will be valid.

Figure~\ref{fig:reffmass} shows that $M_{\ast}$ for all 192 models is
in good agreement with the majority of the observed masses of local
NSCs. In fact, the median remnant mass of
$\log(M_{\ast}/M_{\odot})=6.3$ predicted by the NSD models
(Table~\ref{table:medians}) is in exact
agreement with the median mass of the NSCs shown in the
figure. However, there
are disagreements at the edges of the mass distribution. In particular, the maximum NSD remnant mass is $M_{\ast}
\approx 3\times 10^7$~M$_{\odot}$, more than an order of magnitude
smaller than the maximum observed NSC masses. This supports the view
that some NSCs are likely built up by a variety of processes over time, and
a high-$z$ NSD may only be one of the possible mechanisms. It
is also possible that the mismatch in maximum mass is due to our
limited sampling of the potential parameter space and considering only
a single NSD event during the lifetime of a galaxy. Overall, the
strong agreement between $M_{\ast}$ and the observed NSC masses shows
that the NSDs have the SFRs and lifetimes necessary to explain the old
population of NSCs.


In contrast to the masses, most of the NSD remnants have $\reff$
significantly larger than the observed NSCs. The median predicted
$\reff$ for the NSD remnants is $50$~pc (Table~\ref{table:medians}), an order of magnitude larger than the median of the
observational sample ($\reff = 5$~pc). However,
Fig.~\ref{fig:reffmass} shows that the group of NSD remnants with
$\reff < 10$~pc are in very good agreement with the properties of NSCs
observed in predominantly late-type galaxies (LTGs; red `X's;
\citealt{geo16}) with median $\reff$ and $M_{\ast}$ very similar to
the observations (last line of Table~\ref{table:medians}). These
compact star clusters are predicted to arise from NSDs with the
strongest pc-scale starbursts (e.g., Fig.~\ref{fig:sfrs}). The
scarcity of models at $3\ \mathrm{pc} \la \reff \la 10\ \mathrm{pc}$
is likely due to the choice of model
parameters. Fig.~\ref{fig:reffmbh} shows that the smallest $\reff$
values obtained from the NSD models increase with the central
$\Mbh$ mass. Therefore, NSD remnants with $3\ \mathrm{pc} \la \reff
\la 10\ \mathrm{pc}$ would be produced from models with $\Mbh =
10^{8.5}$~M$_{\odot}$ and $10^{9}$~M$_{\odot}$ and would closely match
the observed NSC sizes in early-type galaxies (ETGs; blue `Y's;
\citealt{geo16}). We conclude that NSDs with the strongest pc-scale
starbursts can produce remnants with $\reff$ similar to those observed in local
galaxies.  

Figure~\ref{fig:reffmass} shows that the majority of NSD remnants have
$\reff \ga 30$~pc which is consistent with only a few observed NSCs hosted
in both ETGs and LTGs. In general, such large $\reff$ values will be a
common outcome from NSD models due to the relatively specific
conditions needed to produce a strong pc-scale starburst
\citep{ball08}. This result implies that only the fraction of high-$z$
NSDs that produce the strongest smaller-scale bursts are a potential origin for
the old population observed in local NSCs. 

\begin{figure}
\includegraphics[width=0.45\textwidth]{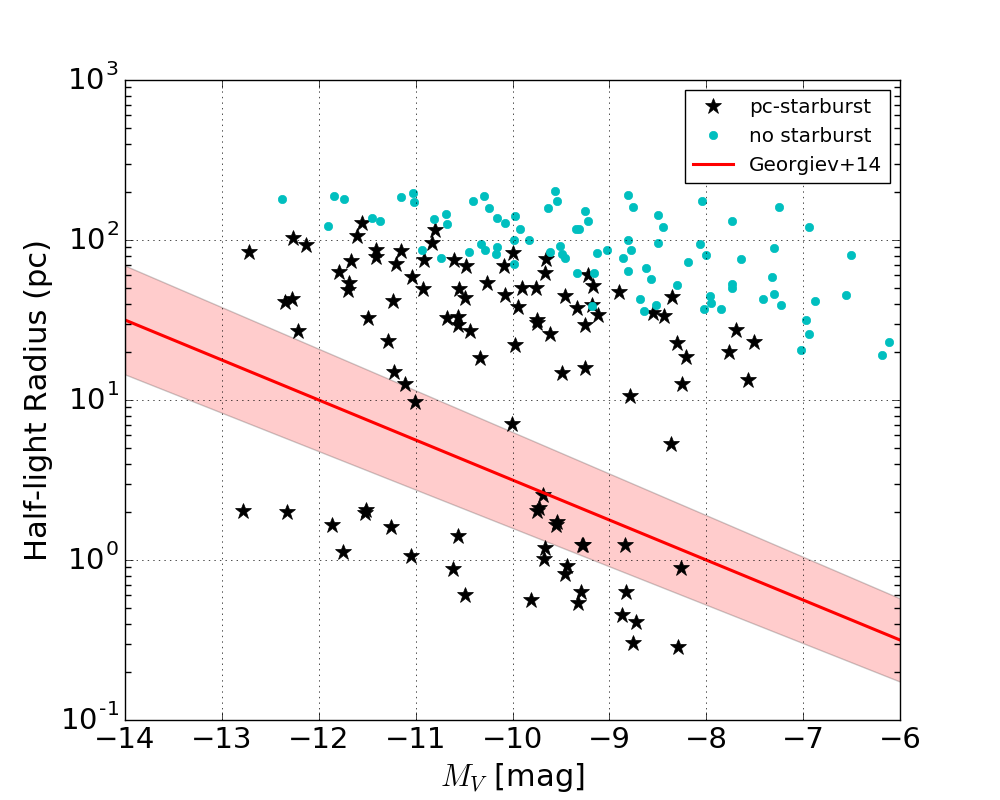}
\centering
\caption{The predicted half-light radius and V-band absolute
  magnitude of the NSD remnants at $z=0.01$, calculated as described
  in Sect.~\ref{sect:calc}. The point colours and styles are the same
  as in Fig.~\ref{fig:reffmass}. The red line and shaded area is the
  size-luminosity relation observed from local NSCs \citep{gb14}.} 
\label{fig:reffmv}
\end{figure}

Another important observational property of local NSCs is a
size-luminosity relation \citep{gb14}. Figure~\ref{fig:reffmv}
compares this observed relation (red line; \citealt{gb14}) with the
predictions of the NSD remnants. The shaded uncertainty region
encompasses the majority of the observed NSCs from \citet{gb14}.
As seen in
Fig.~\ref{fig:reffmass}, most of the NSD remnants have $\reff$
too large for their luminosity, but there is a group of $\approx 40$
remnants that are predicted to lie within or very close to the
observed size-luminosity relationship. 

It is clear that all three sub-groups of NSD remnants follow a
size-luminosity relationship of the form $\log \reff \propto \alpha
M_v$, with $\alpha$ measured to be $\approx -0.11$. This value is roughly
$2\times$ smaller than the observed slope of $\alpha=-0.25\pm 0.02$
found by \citet{gb14}. As the predicted slope does not take into account
the effect of the local galaxy SMBH function on the weighting of the
model parameters, the agreement of the two slopes is considered satisfactory. The values of $M_V$
predicted for the NSD remnants are also well matched with the observed
values (median $M_V= -9.6$; Table~\ref{table:medians}) which is not surprising since the masses are in good
agreement and the luminosities are calculated using standard
mass-to-light ratios \citep{conroy09}.  

\begin{figure}
\includegraphics[width=0.45\textwidth]{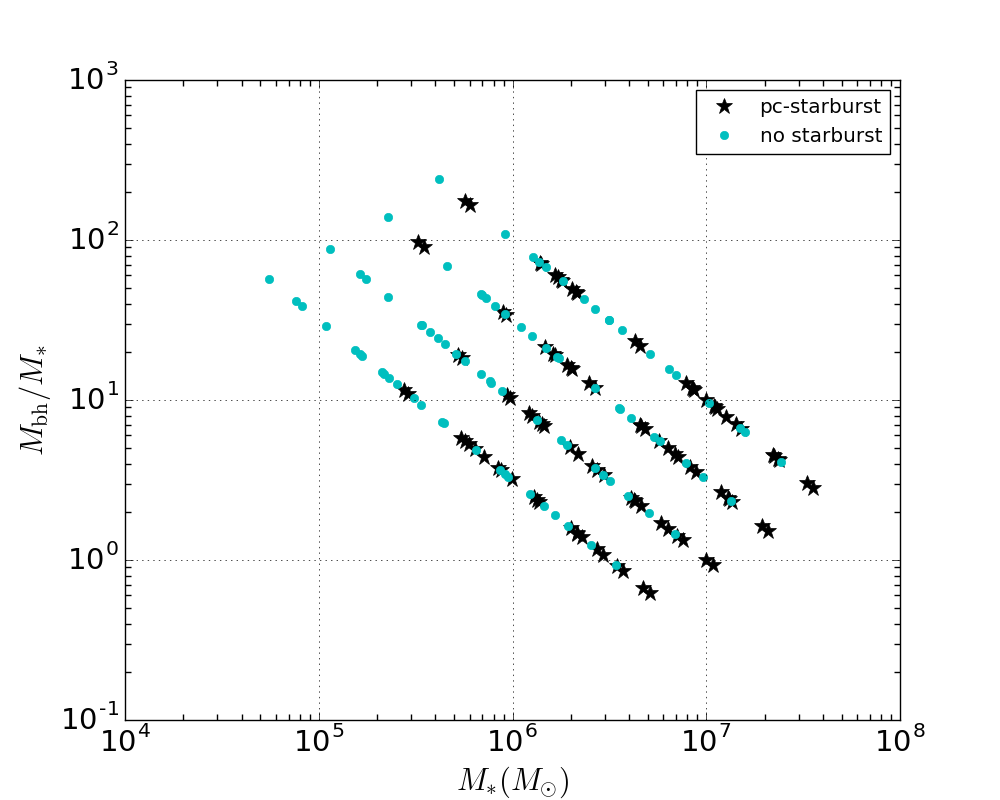}
\centering
\caption{The ratio of the SMBH mass and the
  predicted NSD remnant mass is plotted versus $M_{\ast}$. The point
  colours and styles are the same as in Fig.~\ref{fig:reffmass}. The four
  stripes correspond to the four SMBH masses considered
  (Table~\ref{table:ips}), with the values decreasing from top to
  bottom beginning at $\Mbh=10^8$~M$_{\odot}$. The most massive (and
  therefore most luminous) stellar remnants are produced by the most
  massive SMBHs. The $\Mbh/M_{\ast}$ ratios are consistent with values
most frequently observed in local early-type galaxies as well as a few
late-type galaxies \citep{geo16,nguyen18}.} 
\label{fig:massratio}
\end{figure}

The NSD models require the presence of a SMBH, and, as seen in
Sect.~\ref{sect:calc}, the more massive the central SMBH, the more
likely a compact starburst will develop at high-$z$. It is therefore interesting
to consider the relationship between the observed properties of the predicted
NSD remnant and the SMBH. Figure~\ref{fig:massratio} plots
$\Mbh/M_{\ast}$ as a function of $M_{\ast}$ for the NSD remnants.
The four stripes on the figure correspond to the four SMBH masses used
in the calculations (Table~\ref{table:ips}), with the values
decreasing from $\Mbh=10^8$~M$_{\odot}$ to $10^{6.5}$~M$_{\odot}$ as
one goes from top to bottom in the plot. The most massive and
therefore the most luminous stellar remnants are produced by NSDs
around the most massive SMBHs. However, these SMBHs can also produce
much less massive remnants. Indeed, except at the most massive end,
the NSD theory admits stellar remnants with similar masses and sizes
to local NSCs at all SMBH masses. Measurements of NSC and SMBH masses
from the same galaxies are relatively rare, but current data shows
that ratios $\Mbh/M_{\ast} > 1$ are frequently found in ETGs, but are
observed for only a few LTGs \citep{geo16,nguyen18}. 

As a final sanity check, the WFPC2 colours F606W$-$F814W and F450W$-$F814W
were computed for the stellar remnants and were found to be $0.58$ and
$1.38$, respectively. These values are consistent with the NSC colours
observed by \citet{gb14} and the expectations for a $\sim 10$~Gyr-old
population. 

\section{Discussion \& Summary}
\label{sect:discuss}
The origin of NSCs and their relationship to SMBHs have long been an
interesting question in galaxy formation studies. NSCs are frequently
present in many types of galaxies, and possess a significant older
stellar population, indicating that a large fraction of the cluster
has been in place for several Gyrs. This paper argues that this old
stellar population in NSCs may be the remnant of a NSD that was
present in the galaxy at $z \sim 1$. NSDs may be a common feature
around growing SMBHs during an AGN phase at high-$z$ due to the
large gas fractions within galaxies at this time. Previous work has
shown that they provide a natural origin for the absorbing gas that
obscures most AGNs at $z \sim 1$ \citep{gb17,gb18}. Importantly, NSDs
require the presence of a SMBH, but the size and mass of the remnant
cluster depend only weakly on the SMBH mass (e.g., Fig.~\ref{fig:massratio}). Therefore, the NSD theory
provides an explanation for the complex observed relationship between
NSCs and a central SMBH \citep[e.g.,][]{geo16,nguyen18}. However, as
there are (typically, low-mass) galaxies with a NSC and no SMBH
(e.g., NGC 205; \citealt{nguyen18}) the NSD mechanism proposed here
will not be relevant for all galaxies.

We presented several quantitative comparisons in support of a NSD origin for
the old stellar population of NSCs in galaxies with SMBHs. First, the predicted masses, colours and
luminosities of the remnant stellar cluster correspond nearly exactly
to those observed in NSCs (Figs.~\ref{fig:reffmass} and~\ref{fig:reffmv}). This result only relied on the calculated
equilibrium SFRs in each NSD and the dynamical time of the
discs. Therefore, these predictions are a natural outcome of the
star-forming properties of the disc. If NSDs are common in galaxies at
$z \sim 1$ than remnant clusters will likely exist with these
properties. Secondly, the predicted $\reff$ of the remnant clusters follow a similar
size-luminosity relations as observed in local NSCs
(Fig.~\ref{fig:reffmv}). Finally, the $\Mbh/M_{\ast}$ ratio of the
remnants is also consistent with observations of local NSCs
(Fig.~\ref{fig:massratio}).

Only $\approx 20$\% of the remnants had $\reff$
values consistent with the majority of local NSCs
(Fig.~\ref{fig:reffmass}), with the remainder being too large by about
an order of magnitude, although this fraction depends on the sampling
of the NSD parameter space that was explored (for example, models
with larger $\Mbh$ would result in more remnants with $\reff < 10$~pc). This result
suggests that NSDs are a viable origin for NSCs only when there
is a very powerful pc-scale starburst in the NSD at $z\sim 1$. This scenario is
predicted to be most common in galaxies with a large SMBH mass (Fig.~\ref{fig:reffmbh}). Therefore, NSDs may be a viable origin for NSCs in early-type
galaxies (as supported by the large $\Mbh/M_{\ast}$ ratios) and in
late-type galaxies with large $\Mbh$ masses (e.g., M31). While dynamical
evolution of the stellar distribution in NSD remnants will likely
not significantly impact $\reff$ (Sect.~\ref{sub:dynamics}), VRR will
cause the final remnant to be both rotating and roughly spherical,
consistent with the observed shape of the old populations of NSCs in
some galaxies \citep{seth08b,car15}. 

In summary, it appears that the NSDs with strong
pc-scale starbursts at $z\sim 1$ could be the origin of the old populations of
local NSCs around massive SMBHs. Indeed, the stellar remnants predicted
by this type of NSDs have sizes, colours
and luminosities similar to many nearby NSCs. The stellar remnants are
also expected to be rotating and have evolved into a near spherical
distribution. Testing the predictions presented here will require both new observational and theoretical work. A
clear prediction of our NSD model is that the most massive NSCs will
be associated with the most massive SMBHs. Based on this result, ETGs
may provide the best test cases for the NSD origin hypothesis, as
these galaxies appear
to consistently have the large $\Mbh/M_{\ast}$ ratios predicted by the
theory. Furthermore, the lower gas content in the nuclei of ETGs will
allow for easier charcterization of the old NSC population. Increasing the number of accurate NSC and SMBH mass measurements in
both early- and late-type galaxies will also be important in testing
the theory. However, the most compelling test of the model will be to
search for evidence of NSDs in galaxies at $z \sim 1$, either through
a combination of multi-wavelength measurements
\citep{ball08}, or by direct imaging using instruments carried by the
\textit{James Webb Space Telescope}. Once the incidence of NSDs is
known at high-$z$, then the origin of many nearby NSCs may become much
clearer. Lastly, it remains likely that most nearby NSCs will
have undergone several episodes of growth, possibly driven by
different mechanisms. Therefore, careful observational decomposition
of the star clusters \citep[e.g.,][]{kach18} will continue to be an
important tool in understanding the origin of NSCs.

\section*{Acknowledgments}
The authors thank A.\ Seth and D.\ Nguyen for providing the NSC
observational data used in this paper, and the anonymous referee for
helpful comments that improved the manuscript.








\bsp 

\label{lastpage}
\end{document}